\documentclass[10pt,aps,prl,raggedbottom,longbibliography,reprint,citeautoscript,letterpaper,superscriptaddress]{revtex4-2}
\usepackage[usenames,dvipsnames]{color}
\usepackage{graphicx}
\usepackage{microtype}
\usepackage{booktabs} 
\usepackage{multirow} 
\usepackage{amssymb}

\usepackage[bookmarks=false,colorlinks]{hyperref}
\hypersetup{linkcolor=magenta,citecolor=MidnightBlue,filecolor=Plum,urlcolor=MidnightBlue}
\usepackage[all]{hypcap} 
\usepackage[version=4]{mhchem}
\ExplSyntaxOn
\keys_define:nn { mhchem } 
 {
  arrow-min-length .code:n =
   \cs_set:Npn \__mhchem_arrow_options_minLength:n { {#1} } % default is 2em
 }
\ExplSyntaxOff



\usepackage{xspace}
\newcommand{\supf}{\textcolor{RoyalBlue}{Fig.\ S}}
\newcommand{\supt}{\textcolor{RoyalBlue}{Table} S}

\begin{document}
\mhchemoptions{arrow-min-length=1em}

\title{Heteroanionic Stabilization of Ni$^{1+}$  with Nonplanar Coordination in Layered Nickelates}

\author{Jaye K.\ Harada}
\affiliation{Department of Materials Science and Engineering, Northwestern University, Evanston, Illinois, 60208, USA}

\author{Nenian Charles}
\affiliation{Department of Materials Science and Engineering, Northwestern University, Evanston, Illinois, 60208, USA}

\author{Nathan Z.\ Koocher}
\affiliation{Department of Materials Science and Engineering, Northwestern University, Evanston, Illinois, 60208, USA}

\author{Yiran Wang}
\affiliation{Department of Chemistry, Northwestern University, Evanston, Illinois 60208, USA}

\author{Kenneth R.\ Poeppelmeier}
\affiliation{Department of Chemistry, Northwestern University, Evanston, Illinois 60208, USA}

\author{Danilo Puggioni}
\email{danilo.puggioni@northwestern.edu}
\affiliation{Department of Materials Science and Engineering, Northwestern University, Evanston, Illinois, 60208, USA}

\author{James M.\ Rondinelli}
\email{jrondinelli@northwestern.edu}
\affiliation{Department of Materials Science and Engineering, Northwestern University, Evanston, Illinois, 60208, USA}

\begin{abstract}%
We present electronic structure calculations on layered nickelate oxyfluorides derived from the Ruddlesden-Popper arisotype structure in search of unidentified materials that may host nickelate superconductivity. By performing anion exchange of oxygen with fluorine, we  create two heteroanionic La$_2$NiO$_3$F polymorphs and stabilize Ni$^{1+}$ in 4-coordinate and 5-coordinate square planar and square pyramidal geometries, respectively. We further predict chemical reactions with high thermodynamic driving forces to guide their synthesis. These oxyfluorides are weakly correlated antiferromagnetic insulators and their nonmagnetic phases exhibit quasi-2D Fermi surfaces dominated by Ni $d_{x^2-y^2}$ states, which strikingly resemble  undoped cuprate superconductors. We discuss how the oxyfluoride anion chemistry promotes  Ni-O covalency and single-band character that is more similar to the cuprates than homoanionic infinite-layer nickelates. We use our understanding to propose doping strategies and  layered LaSrNiO$_2$F$_2$ and La$_3$Ni$_2$O$_4$F$_3$ nickelate oxyfluorides with tunable electronic and magnetic structures for experimentation.
\end{abstract}

\date{\today}

\maketitle
 \begin{figure*}
\centering
\includegraphics[width=0.7\columnwidth]{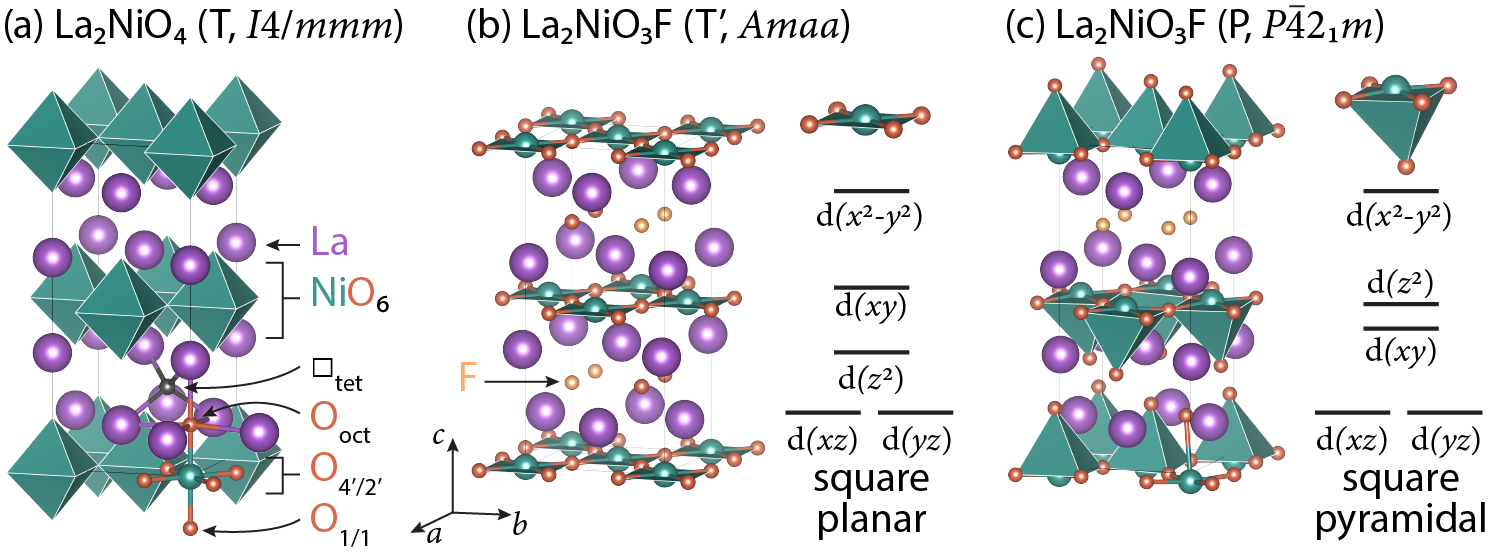}
\caption{(a) Aristotype Ruddlesden-Popper La$_2$NiO$_4$ oxide nickelate with the T-type structure and derived La$_2$NiO$_3$F polymorphs with the (b) T$^\prime$ structure  %(space group $Amaa$) 
and (c) P structure. 
The corresponding representative crystal field splitting energy levels are 
show for both the square planar and square pyramidal Ni coordination geometries.
} 
\label{fig:structure}
\end{figure*}

%\emph{Introduction.}---
The electronic structure of complex transition metal compounds can be designed to exhibit a myriad of correlated physics  and functional phenomena \cite{Zhang:Song:Madsen:Thermoelec/2016, Wu/Chu_et_al:1987, PhysRevLett.110.156403, Bockris/Otagawa:1984, Imada/Fujimori/Tokura:1998, Ohta/Kenji:2008}.
$R$NiO$_3$ nickelates are an exemplary family with this  tunability \cite{Zhang_2022}. 
They exhibit metal-to-insulator, paramagnetic-to-antiferromagnetic, and bond/charge-ordering transitions with critical temperatures governed by choice of the rare earth $R$ cation \cite{Chaloupka/Khaliullin:2008, varignon2017complete, middey2016physics}, which inductively modifies the Ni $d$--$p$ orbital interactions \cite{GoodenoughChengRe:2010}. %
Further tailoring may be obtained  with epitaxial strain, heterostructuring, or light to modulate and even suppress these transitions \cite{PhysRevB.106.165104}.
More recently, the electronic-orbital dependencies on the Ni cation coordination has been exploited in search of novel nickelate superconductors. 
The guiding discovery approach has been to mimic the electronic structure of  superconducting cuprate oxides and oxyhalides, where Cu exhibits planar or square pyramidal coordination \cite{Cu_sqpl,Cu_sqpyr,Cu_halides}, by modifying the atomic scale structure to control the orbital occupations \cite{Anisimov/Bukhvalov/Rice:1999,Hansman:2010,HanMills2010, PhysRevLett.102.046405,Hanghui/Kumah/Disa/Ahn/Beigi:2013,Yee2015}. 
Most attempts in 3D nickelates have led to minimal orbital polarizations and the absence of superconductivity. 

Nickelate superconductivity remained unsuccessful until the recent synthesis of two-dimensional (2D) thin films of Sr-doped NdNiO$_2$ and PrNiO$_2$ \cite{Li:2020,Danfeng:2019,PhysRevMaterials.4.121801,Cano:2020,Norman:2020,Bernardini:2022,Goodge:2023}.
These were obtained through a process of thin-film deposition followed by soft-chemistry methods  \cite{Hayward2003}, utilizing a strong topotactic metal hydride reduction reaction that simultaneously achieves both a planar 2D atomic structure and electrochemically drives Ni into the difficult to obtain $+1$ oxidation state \cite{Anisimov/Bukhvalov/Rice:1999} with a formal $t_{2g}^6e_g^3\, (S=1/2$) electronic configuration matching Cu$^{2+}$ in the isostructural infinite-layer cuprate CaCuO$_2$. 
Nonetheless, the observed maximum superconducting critical temperature ($T_c\sim15$\,K) is well below that of high-$T_c$ cuprates ($T_c\sim133$\,K) \cite{Scilling:1993}, and the nickelate transport properties strongly depend on sample quality \cite{Li:2020,PhysRevLett.125.147003, Osada:2020}. 
This sensitivity likely arises from the aggressive reduction reaction, which although creating and replicating the 2D cuprate topology, also leads to unintentional hydride incorporation to form a mixed-anion nickelate oxyhydride  with heteroleptic NiO$_4$H$_2$ octahedra \cite{Ding_2023}. 
These  findings suggest H is an important ingredient to both realizing the 2D structure and superconductivity---the materials may be better described as heteroanionic infinite-layer nickelates  comprising more than one anion.

Here we propose the intentional creation of heteroanionic nickelates as an alternative strategy to realize analogous cuprate physics, and as we emphasize below, essential materials chemistry, by starting from quasi-2D bulk oxides and exploiting  multi-anion rather than cation effects \cite{Kitamine2020,Hirayama2020,Bernardini:2021,PhysRevMaterials.6.114404}. 
Consider the $R_{n+1}$Ni$_n$O$_{3n+1}$ Ruddlesden-Popper (RP) structure, which exhibits alternating $n$ perovksite planes of corner-connected NiO$_6$ octahedra interleaved between rock-salt $R$O sheets. 
The ideal $n=1$ RP oxide with the K$_2$NiF$_4$- or T-type structure (space group $I4/mmm)$ hosts  Ni$^{2+}$ ($t_{2g}^6e_g^2$, $S=1$) and is distinct from other layered nickelates %, e.g.,  $R_{n+1}$Ni$_n$O$_{2n+2}$ 
\cite{Zhang2017f,Pan_2021}. 
It  may be written as $(R^{3+})_2{\,}^{~2}_{\infty}[\textrm{Ni}\textrm{O}_{4^\prime /2^\prime}\textrm{O}_{2/1}]^{6-}$ \cite{JENSEN1989105}, revealing that there are infinite layers of octahedra comprising 4 linear-bridging and 2 terminal oxide anions (\autoref{fig:structure}a). 
Since fluorine prefers to occupy terminal sites in RP oxyfluorides \cite{Harada_2019}, equal exchange of oxygen with the lower valence fluoride anion reduces the metal to Ni$^{1+}$ while splitting the apical sites: $(R^{3+})_2{}^{~2}_{\infty}[\textrm{Ni}\textrm{O}_{4^\prime /2^\prime}\textrm{O}_{1/1}\textrm{F}_{1/1}]^{6-} = R_2\textrm{NiO}_3\textrm{F}$. 
Hence, the square-planar geometric requirement proposed by Anisimov \cite{Anisimov/Bukhvalov/Rice:1999}, and pursued over the last three decades of nickelate research to stabilize Ni$^{1+}$, can be inherently circumvented in oxyfluorides. 
Yet, there remains ambiguity in whether 
the octahedral geometry persists or transforms to  square planar or square pyramidal coordination, and in doing so, which sites the displaced ligands occupy.
We  examine with density functional theory (DFT) the variety of structure types and local environments that RP-derived oxyfluoride nickelates adopts in search of candidate bulk heteroanionic nickelate superconductors.
We create two polymorphs differing by their local Ni coordination environment and position of the fluoride anions relative to the oxygen anions. 
The first phase of La$_2$NiO$_3$F exhibits square-planar coordinated Ni cations (\autoref{fig:structure}b), and  we refer to  as  T$^\prime$ following the convention for this hettotype (\supt{1} of the Supporting Information \cite{Supp}).
The structure is obtained from a $\sqrt{2}\times\sqrt{2}\times1$ transformation of the RP $I4/mmm$ structure with ordered 
terminal anion vacancies, i.e., $(\textrm{La})_2{\,}^{~2}_{\infty}[\textrm{Ni}\textrm{O}_{4^\prime /2^\prime}\square_{2/1}]$. 
Then, interstitial tetrahedral sites in the (001) plane between the La cations are decorated with [100] anion stripes that alternate between O and F along [010] to produce $(\textrm{La}_2\textrm{O}_{4^{\prime\prime}/4}\textrm{F}_{4^{\prime\prime}/4}){\,}^{~2}_{\infty}[\textrm{Ni}\textrm{O}_{4^\prime /2^\prime}]$ or the La$_2$NiO$_3$F stoichiometry \cite{Wissel2022}.
Adjacent planes of these interstitial anions are translated relative to each other by 1/2$b$. 
We then performed a full relaxation of the structure with DFT 
\cite{Supp}, and found that the equilibrium geometry exhibits anti-polar La displacements along the [110] direction which reduces the crystal symmetry to orthorhombic $Amaa$. 
Previously, a tetragonal structure with a different O-F arrangement and $I\bar{4}m2$ symmetry was studied \cite{Bernardini:2021}; however, the calculated energy difference between the $I\bar{4}m2$ and $Amaa$ phases  is $\sim$100\,meV/f.u.\ favoring the latter as more stable.

Next, we examined  La$_2$NiO$_3$F with square-pyramidal Ni coordination, which we refer to as the P structure (\supt{1}).
It is a derivative of  the ordered $n=1$ RP heteroanionic oxyfluoride Sr$_2$FeO$_3$F  (space group $P4/nmm$) \cite{app2010206}. 
In contrast to the T-type structure,  Sr$_2$FeO$_3$F exhibits \emph{trans} ordering of the terminal O/F anions at the apical octahedral sites. 
This results in a strong distortion of the %FeO$_5$F octahedron 
$^{~2}_{\infty}[\textrm{Fe}\textrm{O}_{4^\prime /2^\prime}\textrm{O}_{1/1}\textrm{F}_{1/1}]$. 
The F anion significantly displaces away from Fe along the \emph{trans} bond into the Sr  layers but it does not occupy the tetrahedral interstitial site, i.e.,  $(\textrm{Sr}_2\textrm{F}_{1/1}){\,}^{~2}_{\infty}[\textrm{Fe}\textrm{O}_{4^\prime /2^\prime}\textrm{O}_{1/1}]$.
We fully relaxed this structure and several distorted variants of this derivative  and found that unlike Sr$_2$FeO$_3$F, the fluoride ions always move from the octahedral (oct) to tetrahedral (tet) interstitial site to form NiO$_5$ square pyramidal polyhedra (\autoref{fig:structure}c). 
In both T$^\prime$ and P  polymorphs the fluorine ions prefer to occupy the interstitial tetrahedral sites (\autoref{fig:structure}b,c).
The lowest energy P structure  is noncentrosymmetric with tetragonal $P\bar{4}2_1m$ symmetry and  exhibits alternating $a^-b^0c^0$ and $a^0b^-c^0$ Glazer tilts \cite{Glazer:1972} in adjacent layers along the [001] direction. 
In layered cuprate superconductors,
the copper--apical-oxygen distance (Cu--O$_{ap}$) correlates with the maximal $T_c$ \cite{Slezak:2008,PhysRevLett.87.047003,PhysRevB.45.10647}; in the P structure, the Ni-O$_{ap}$  distance is 2.62\,\AA, which is intermediate between YBa$_2$Cu$_3$O$_{6.9}$ ($\sim2.28$\,\AA\, \cite{Voronin:2000}) and HgBa$_2$CuO$_6$ ($\sim2.80$\,\AA\, \cite{EVAntipov_2002}).

We further confirmed the  dynamical stability of both phases (\supf{1}). 
Although both polymorphs are dynamically stable, the P structure is $\approx570$\,meV higher in energy than the T$^\prime$ phase. 
Recent synthesis of La$_2$NiO$_3$F and Pr$_2$NiO$_3$F using solid-solid reactions with topochemical hydride-based defluorination supports this finding, which reported structures with square planar NiO$_4$  \cite{Wissel2020,Wissel2022}.
Magnetic measurements on these oxyfluoride powders found spin-glass behavior similar to the infinite layer oxide nickelates, while  DFT calculations predict a strongly antiferromagnetic phase \cite{Bernardini:2021}. 
Such discrepancies may arise from inhomogeneity in the fluoride concentration, which causes variability in the local Ni coordination, valence, and overall hole concentration. 

\begin{figure} 
\centering
\includegraphics[width=0.93\columnwidth,clip]{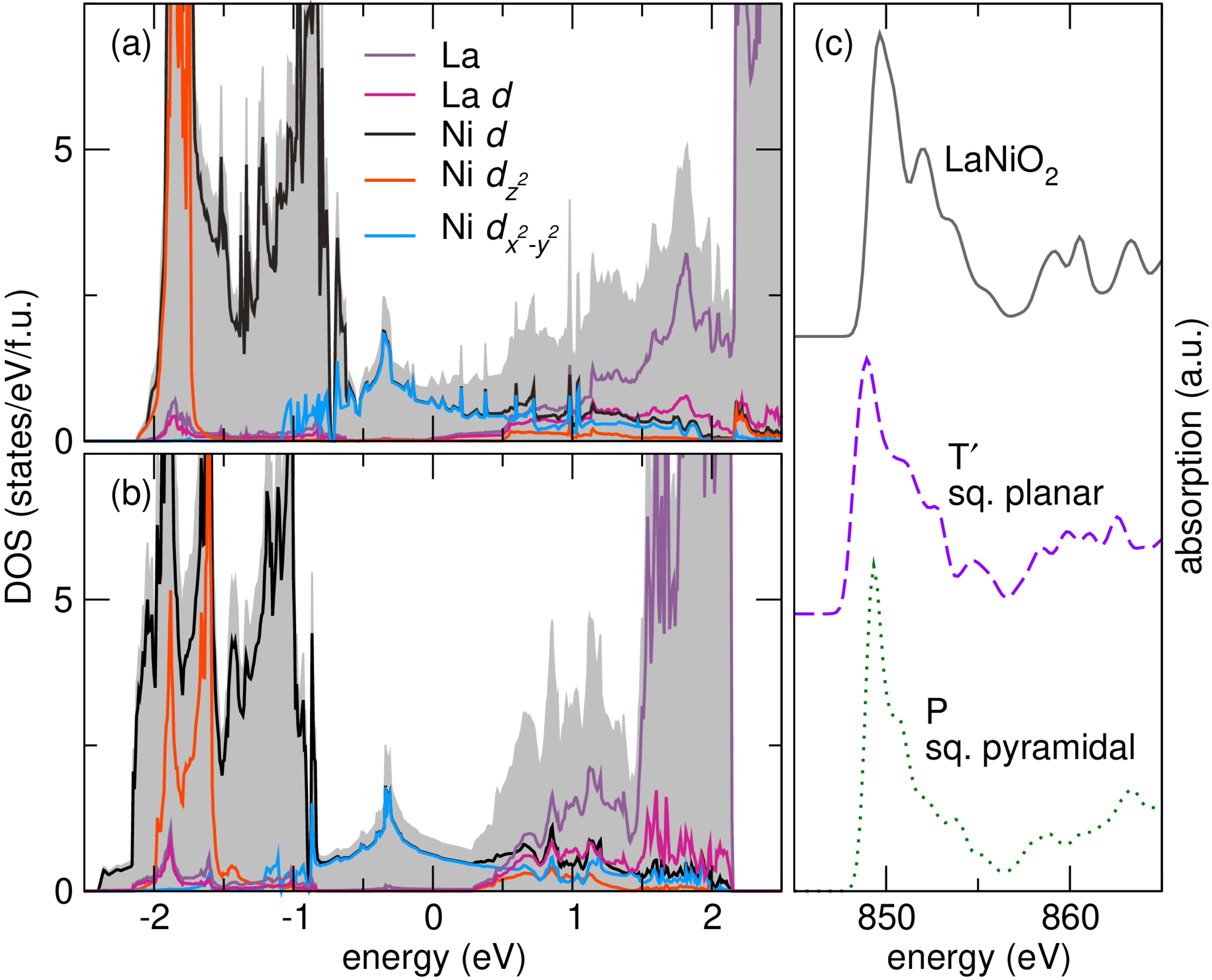}
\caption{Projected density of states (DOS) of La$_2$NiO$_3$F with the (a) T$^\prime$ and (b) P structures with E$_F=0$\,eV. (c) Simulated Ni L-edge XAS spectra of relevant nickelates.
} 
\label{fig:nsp_dos}
\end{figure}

These complexities may be overcome by identifying a synthetic reaction that provides a high thermodynamic driving force for phase formation \cite{Sun_etal:2021}. 
To that end, we calculate the reaction chemical potentials ($\Delta \mu_{\textrm{rxn}}$) for several solid-solid, solid-liquid, and solid-gas reactions to assess which provides the highest driving force for the P structure %La$_2$NiO$_3$F with square pyramids 
(\supt{2}). 
The solid-state reaction, i.e.,  \ce{LaNiO2 + LaOF -> La2NiO3F},  which does not change the Ni oxidation state, does not provide a sufficient driving force for either the %square pyramidal (P)  
P or lower-energy T$^\prime$ 
%square planar (T$^\prime$) 
phases ($\Delta \mu_{\textrm{rxn}} = 0.82$\,eV/f.u.\ 
 and $0.24$\,eV/f.u., respectively). 
Thus, we pursued distinct Ni redox pathways and found that the most promising reactions to produce the %square pyramidal 
P polymorph were a solid-solid reaction with sodium hydride 
and a reaction with hydrazine in aqueous HF(aq):
\begin{eqnarray*} 
\ce{2La2NiO3F2 + 2NaH &-> 2La2NiO3F + 2NaF + H2} \\ 
%($\Delta \mu_{\textrm{rxn}}  = -1.09$\,eV/f.u.), which we know produces the square planar variant,\cite{Wissel2020} 
\ce{4La2NiO4 + 4HF + N2H4 &-> 4La2NiO3F + 4H2O + N2} 
\end{eqnarray*} 
with $\Delta \mu_{\textrm{rxn}}  = -1.09$\,eV/f.u. and -0.33\,eV/f.u., respectively. %
Because hydrothermal reactions can produce metastable phases and have tunable pH, temperature, and ion concentrations, which enable high polymorphic selectively of heteroanionic oxyfluoride and oxychalcogenide products \cite{Chang2014,Walters2021}, we believe this is a viable route to produce the P structure. 
Phase purity of it can be ascertained both with X-ray diffraction and second harmonic generation measurements.
%%

%\emph{Nonmagnetic electronic properties.}---
Next, we examine the projected density of states (DOS) for the nonmagnetic phases of both  the T$^\prime$ and 
%square pyramidal 
P 
%La$_2$NiO$_3$F 
polymorphs (\autoref{fig:nsp_dos}a,b).
\supf{2} shows the DOS over a wider energy range with more orbital projections. 
%and is consistent with Ref.\ \cite{Bernardini:2021}.
%
Both nonmagnetic phases are metals and the Fermi level (E$_F$) is dominated by the 
Ni $3d_{x^2-y^2}$ states. 
The La 5$d$ states located nearby at higher energy \cite{Bernardini:2021}, 
and they self-dope the T$^\prime$ phase as in homoanionic nickelates but  are fully empty in the P phase.
Owing to the different Ni coordination environments, the Ni $3d$ orbitals crystal field splitting energy diagrams are distinct  (\autoref{fig:structure}). 
This is clearly seen through the position of the $3d_{z^2}$ orbital, which comprises the bottom of the nominal Ni $3d$-derived valence band around $\sim-1.9$\,eV in the T$^\prime$ structure compared to the P structure at $\sim-1.7$\,eV.
The O and F $2p$ states are separated from the Ni $3d$ states by a $\sim$1\,eV electronic band gap (\supf{3}). 
First, the F $2p$ states are located below $\sim$-6\,eV and exhibit a different degree of localization in the two phases.
Those states are more localized in the T$^\prime$ crystal structure. This is due to the 1D striped arrangement of the F anions in the structure in contrast to the 2D net in the P structure. 
Second, the O $2p$ states fall into two separate energy windows: 
approximately 
[-8.2,-3.2] and [-2,-0.6]\,eV for the T$^\prime$ phase and 
[-8.8,-3.5] and [-2.1,-0.9]\,eV for the P phase (\supf{2} and \supf{3}). 
This distribution alters the O($2p$)-Ni($3d$) hybridization and has a profound effect on covalency (infra vide).
%%

%\begin{figure*}
\begin{figure}
\centering
\includegraphics[width=0.49\textwidth,clip]{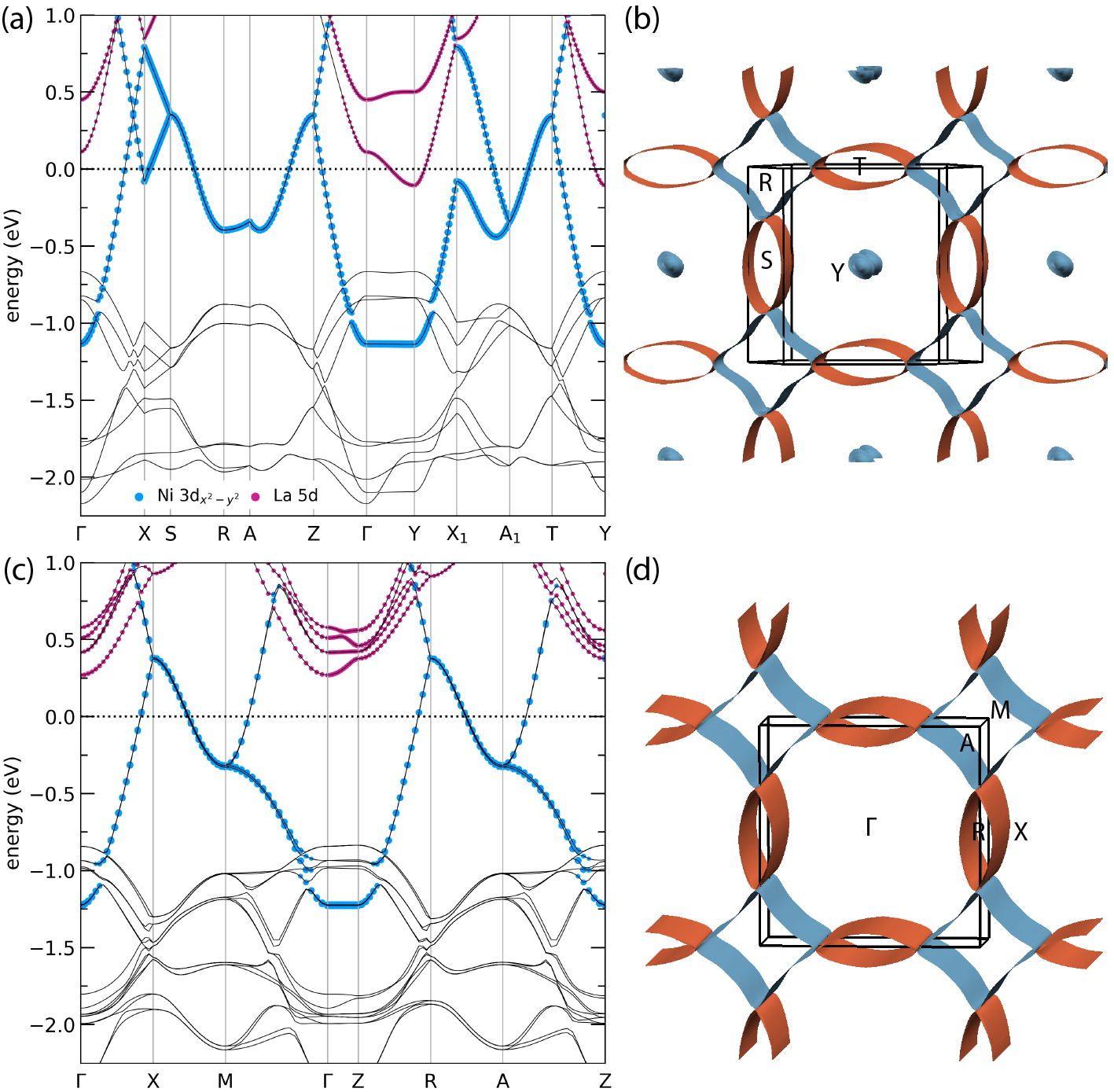}
\caption{Band structures (a,c) and Fermi surfaces (b,d) of %square planar and square pyramidal
T$^\prime$ and P La$_2$NiO$_3$F polymoprphs, respectively. Both Fermi surfaces exhibit electron (light blue) and hole (orange) pockets.} 
\label{fig:nsp_bandstruct_fermi}
%\end{figure*}
\end{figure}

Our analysis %of the electronic structure 
of both La$_2$NiO$_3$F polymorphs %, in agreement with Ref.\,\onlinecite{Bernardini:2021}, 
suggests that Ni is in the 1+ oxidation state ($3d^9$) and is isoelectronic to Cu$^{2+}$. 
To assess %further confirm 
the oxidation state, we computed the Ni L-edge X-ray absorption spectrum (XAS) \cite{Supp}. 
\autoref{fig:nsp_dos}c shows the simulated spectra for T$^\prime$ and P La$_2$NiO$_3$F and the infinite-layer nickelate LaNiO$_2$.  All simulated spectra occur at the similar energy, starting at approximately 846 eV, suggesting that each compound comprises Ni cations with the same valence.  The lineshapes of the spectra are also similar.  Our calculated LaNiO$_2$ spectrum exhibits two peaks in a 2\,eV range from $\sim850$ to $\sim852$\,eV.  Both the T$^\prime$ and P structures 
have two peaks in this range such that  %the 850-852\,eV range, and
the low-energy  L-edge spectra are similar to LaNiO$_2$.     
Next, we validate our calculated LaNiO$_2$ spectra by comparing to experiment upon aligning the first peak of each spectra \cite{Karsai:2018}.  The experimental spectrum of LaNiO$_2$ 
shows two peaks in about a 2\,eV range \cite{Hepting2020}, approximately from 852 to 854\,eV, which is similar to the two peaks obtained in a 2\,eV peak range from our model.       
This agreement %data (theory-theory and theory-experiment spectra comparisons) 
supports the assignment of Ni$^{1+}$ in two distinct coordination environments. 

\autoref{fig:nsp_bandstruct_fermi}a,b shows the nonspin-polarized band structure and Fermi surface for T$^\prime$ La$_2$NiO$_3$F. 
%square planar crystal structure, respectively. 
%
The only bands crossing E$_F$ originate from the Ni $3d_{x^2-y^2}$ and La $5d_{z^2}$ orbitals. Moving along the $S$-$Z$ and $A_1$-$T$ directions, the Ni $3d_{x^2-y^2}$ band is doubly degenerate. 
Then it splits and forms two single degenerate bonding and antibonding bands along the $\Gamma$-$S$, $Z$-$A_1$, and $T$-$Y$ trajectories. 
The Fermi surface is quasi-2D and composed of two hole pockets at $S$ and $T$ and one electron pocket at $R$ with Ni $3d_{x^2-y^2}$ character. 
Moreover, there is an electron knob at $Y$ from the La $5d_{z^2}$ band. 
The corresponding electronic structures for the P structure 
are shown in \autoref{fig:nsp_bandstruct_fermi}c,d. 
As in the T$^\prime$ structure, the highly dispersive Ni $3d_{x^2-y^2}$ band exhibits 2D character and produces a Fermi surface with hole and electron pockets centered around the $X$ and $M$ points of the Brillouin zone, respectively. 
The major electronic difference between the square pyramidal P structure and the square planar 
T$^\prime$ structure is that the position of the La $5d_{z^2}$ band is well above the $\Gamma$ point ($\sim0.25$\,eV) and does not cut E$_F$ in the P structure (cf.\ \autoref{fig:nsp_bandstruct_fermi}a,c).

Interestingly, both  Fermi surfaces exhibit reduced O($2p$)-Ni($3d$) hybridization as the oxygen states are shifted to lower energy and more strongly mix with orbitals without $d_{x^2-y^2}$ character.
This is in contrast to that found in the high-$T_c$ cuprates where Zhang-Rice singlets and orbital symmetry are important \cite{Lee_Pickett:2004, PhysRevB.37.3759, Norman:2020,Bernardini:2021}.
Furthermore, our computed Fermi surfaces are distinct from those of the 
infinite-layer  nickelate LaNiO$_2$ \cite{Norman:2020} 
and a nickelate oxyfluoride phase studied in Ref.\,\onlinecite{Bernardini:2021}. 
Apart from the different space group (Brillouin zone) and lattice periodicity (band folding), we find nested  electron and hole 
pockets with Ni $3d_{x^2-y^2}$ character touch each other. 
This suggests that, in both polymorphs, the Ni band $3d_{x^2-y^2}$ is exactly half-filled, as expected in an undoped scenario where the oxidation state of Ni is exactly 1+.

%\emph{Magnetic properties.}---
We also find that both %square planar and square pyramidal 
T$^\prime$ and P polymorphs exhibit G-type antiferromagnetic (AFM) order with antiferromagnetic (001) in-plane nearest-neighbor Ni superexchange. 
The G-type magnetic ordering is 38\,meV and 23\,meV lower in energy 
than the nonmagnetic model for the  
%square planar and square pyramidal
T$^\prime$ and P structures, %of La$_2$NiO$_3$F
respectively. 
The computed local magnetic moments are $\sim$0.60\,$\mu_B$ 
%and 0.57\,$\mu_B$ per Ni site in the %square planar and square pyramidal 
for both polymorphs, %structures %of La$_2$NiO$_3$F
%, respectively, 
further confirming Ni$^{1+}$. % of the Ni cation. 
%
%All attempts to stabilize ferromagnetic order failed, resulting in a nonmagnetic phase, similar to other layered nickelate simulations 
%as already observed in similar square planar nickelate oxyfluorides.
%\cite{Bernardini:2021}.
%%
%
We further modeled  the exchange interactions for Ni$^{1+}$ ($S=1/2$)
using a Heisenberg spin Hamiltonian to obtain the in-plane exchange $J_{in}=\,-38$\,meV and $-23$\,meV for the  %square planar and square pyramidal
T$^\prime$ and P structures \cite{Supp}, respectively. %
%\footnote{We use $H=E_{nm}-\sum_{i,j}J_{i,j}\bar{S_i}\cdot\bar{S_j}$, where $E_{nm}$ is the nonmagnetic part of the total energy and $J$ are the exchange parameters ($J>$0 for ferromagnetic exchange and $J<$0 for the antiferromagnetic case). Independent of the polymorph, each Ni site has  four in-plane and eight out-of-plane nearest neighbors with exchange integrals indicated as $J_{in}$ and $J_{out}$, respectively; we write the total G-type AFM energy as $E_{AFG}=E_{nm}+4J_{in}S^2+8J_{out}S^2$, which becomes $E_{AFG}=E_{nm}+4J_{in}S^2$ because the sum over the out-of-plane spin coupling in the G-type AFM cancels. We use this expression to  calculate the in-plane exchange integrals as $J_{in}=(E_{AFG}-E_{nm})/4S^2$.
%}.
%
%In this particular situation, these exchange interactions are none other than the energy difference between the G-type AFM and the nonmagnetic phase.)
%
The AFM in-plane exchange of  La$_2$NiO$_3$F  is strongly reduced compared to that of cuprate compounds such as  La$_2$CuO$_4$ and CaCuO$_2$ ($J=120$\,meV) \cite{Braicovich:2009}.

\begin{figure}
\centering
\includegraphics[width=0.4\textwidth,clip]{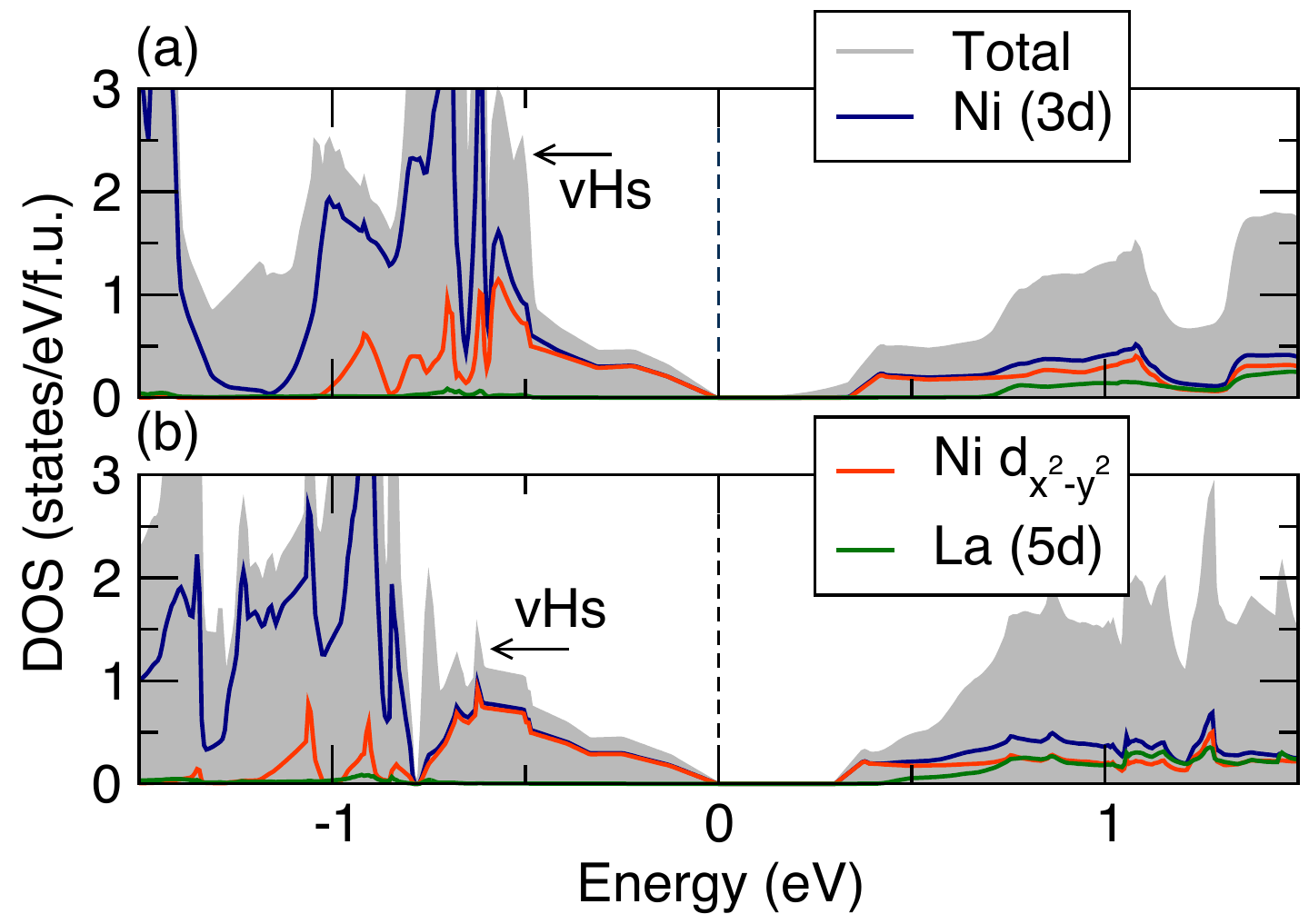}%{figures/Figure4.pdf}
\caption{The projected density of states (DOS) of the AFM G-type (a) 
T$^\prime$
%square planar 
and (b) P polymorphs of
%the square pyramidal 
La$_2$NiO$_3$F. A shift of the Fermi level to the van Hove singularities (vHs), indicated by the arrows, corresponds  to  doping of 0.25\,hole/f.u.\ and 0.33\,hole/f.u\ for the T$^\prime$ and P phases,  respectively. 
}
\label{fig:U}
\end{figure}

Both magnetically ordered AFM-G T$^\prime$ and P polymorphs 
are insulating (\autoref{fig:U}a,b) with DFT electronic band gaps of  
$E_g=150$\,meV  and 300\,meV, 
respectively, which makes them  Slater insulators and conducive to ambipolar doping %.
\cite{Supp}.
Undoped cuprates are described as AFM Mott insulators \cite{RevModPhys.78.17} and the van Hove singularities (vHs) in their electronic structure are used to estimate optimal doping for maximizing $T_c$ \cite{Pickett:1998}. In the DOS shown in  \autoref{fig:U}, the vHs are located at 0.51\,eV and 0.62\,eV below E$_F$, which gives optimal dopings at 0.25\,hole/f.u. and 0.33\,hole/f.u for the T$^\prime$ and P polymorphs, respectively, where minima in the Seebeck coefficients are also expected as in cuprates \cite{PhysRevX.12.011037}. 
Although these hole concentrations are higher than reported for cuprates ($\sim0.16$\,hole/f.u.) \cite{Doiron:2007} and layered square-planar nickelate oxides ($\sim0.2$\,hole/f.u.) \cite{Pan_2021}, they should be readily accessible through cation substitution, e.g., alkaline-earth metals, or through anion insertion with excess oxygen onto unoccupied interstitial sites.
%to maintain the Ni$^{1+}$ coordination geometry.

%%
We also examined the electronic structures of two other Ni$^{1+}$ oxyfluorides with O:F anion ratios: LaSrNiO$_2$F$_2$ and La$_3$Ni$_2$O$_4$F$_3$, which have square planar [NiO$_{4^\prime /2^\prime}$]$^{7-}$ and square pyramidal [NiO$_{4^\prime /2^\prime}$F$_{1/1}$]$^{8-}$ coordinations, respectively (\supt{3}). 
They are also narrow gap AFM Slater insulators \cite{Supp}.
All oxyfluorides examined exhibit 2D electronic structures like that recently predicted for infinite-layer nickelates with hydrogen incorporation \cite{oxyhydride:2023}, where H$^-$ anions were found to occupy apical Ni coordination sites, because the absence of $\pi$ symmetry in the hydride $1s$ orbital makes it a chemical ``scissor'' that blocks interlayer electron hopping. 
In contrast, the 2D Fermiology of the oxyfluorides and hole dopability is due to the distance between the NiO$_4$ plane, which  is $c/2$  ($<\,c/2$) in the T$^\prime$ (P) structure, and the LaO layers. The interlayer distance is large because it is enhanced by interstitial F occupation. 
These oxyfluoride crystal chemistry principles may be used in the same manner as employed by Goodenough and Manthiram to understand how interlayer stresses control electronic compensation and superconductivity in layered cuprates \cite{Goodenough1990}.
%

%The cuprates exhibit both square pyramidal and square planar coordination like %
The covalency and bond anisotropy in cuprates is fundamental to their large  superexchange strengths and high $T_c$ \cite{Walters_2009}.
The RP-derived  oxyfluorides possess this added feature of exhibiting more overall covalent character (less ionicity) compared to the homonanionic layered nickelate oxides.
Despite the addition of more electronegative F$^-$, which reduces the in-plane AFM exchange, 
the fluoride states at lower energy constrain and shift the active  oxygen band centers closer to the $d$ orbitals. 
The reduced energy separation, created through an inductive effective   \cite{Balachandran_2014}, strengthens the formation of the 
low energy antibonding $d$-orbital derived states (\supf{3)}.
We find the oxyfluorides are more covalent than the prototype LaNiO$_2$ owing to increased O($2p$)-Ni($3d$) overlap at the expense of reduced orbital bandwidth from the multi-anion effect. % producing the 2D Fermi surface.
The response is analogous to interlayer charge transfer found in layered cuprates mediated by oxygen displacements \cite{egami1996physical}, where out-of-plane electron transfer increases the in-plane ligand hole density \cite{PhysRevLett.124.207004}. 
We quantify this effect in terms of percentage ionicity 
%following Pauling's approach adapted for complex multianion compounds 
\cite{Charles:2016}.
The ionicities of the nickelate oxyfluorides are  similar and closer to the cuprates than the homoanionic nickelates: % (\supt{4}):
49.1\,\% for La$_2$NiO$_3$F, % polymorphs have ionicities of 49.1\,\%,
51.2\,\% for LaSrNiO$_2$F$_2$, and %$ is 51.2\,\%, and La$_3$Ni$_2$O$_4$F$_3$ 
48.8\,\% for La$_3$Ni$_2$O$_4$F$_3$ versus %is 48.8\,\%. 
the undoped superconductors YBCO, CaCuO$_2$, and LaNiO$_2$ with ionicities of 49.4\,\%, 52.0\,\%, and 57.0\,\%, respectively. 
Additional similarities and differences are in \supt{4}.
%
 
%
%\section{Conclusion}
%\emph{Conclusion.}--
In summary, we investigated the structural, electronic and magnetic properties of several Ni$^{1+}$ oxyfluorides and showed that they have more similarities to cuprates than the infinite layer nickelate oxides.
As undoped compounds, they are weakly correlated antiferromagnetic insulators exhibiting near or complete isolation of the $d_{x^2-y^2}$ orbital enabled by F$^-$. 
In T$^\prime$ La$_2$NiO$_3$F, we found the $R$-derived $5d$ state is near the Fermi level, as in other (hydrogen incorporated) infinite layer nickelates, but that state is absent in the P polymorph.
This makes oxyfluorides an attractive platform to discern contributions from multiband physics to the superconductivity mechanism and whether those states limit access to higher temperature scales.
We  further identified that bond covalency may be an essential component neglected in prior searches. 
This finding should be interpreted through more detailed studies of electron correlation effects to better understanding the role played by 
spin fluctuations at finite temperatures. 
%
%Our results demonstrate the potential for nickelate superconductivity in a wide variety of oxyfluoride phases.
%
Incorporating doping of  nickelate oxyfluorides  using  hydrothermal reactions and/or topotactic fluorination of thin films, which have been shown to produce metastable phases and compliments current thin-film approaches based on heteroepitiaxy or superlattice formation \cite{PhysRevB.104.165137,Goodge_2023}, are needed.
Given that oxyfluorides  adopt a wide variety of structure types, including some that are unique from complex oxides \cite{Tsujimoto2012a,Kobayashi2018,Harada/Pop/Rondo:2019,kageyama2018expanding}, we conjecture there are several unidentified superconducting nickelate oxyfluoride candidates to be found. 
Similarly, nickel oxyhalides with Cl$^-$ or Br$^-$  are other exciting materials families.

\begin{acknowledgments}
Electronic structure studies were supported by the National Science Foundation (NSF) through award DMR-2011208 while the synthesis science efforts were supported by the NSF's MRSEC program (DMR-1720319) at the Materials Research Center of Northwestern University. Calculations were performed using the Department of Defense High Performance Computing Modernization Program (DOD-HPCMP) and the Carbon cluster at the Center for Nanoscale Materials, a U.S. Department of Energy Office of Science User Facility, supported by the U.S.\ DOE, Office of Basic Energy Sciences, under Contract No. DE-AC02-06CH11357.
\end{acknowledgments}

\bibliography{thesis_refs,nickelate_superconductors,references}

\end{document}